\documentclass[12pt]{article}
\usepackage{amsmath}
\usepackage{graphicx}
\usepackage{natbib}
\usepackage{url} 
\usepackage{relsize}
\usepackage{multirow}

\newcommand{\blind}{0}

\begin{document}

\def\spacingset#1{\renewcommand{\baselinestretch}%
{#1}\small\normalsize} \spacingset{1}


\if0\blind
\title{A goodness-of-fit diagnostic for count data derived from half-normal plots with a simulated envelope}
  \author{Darshana Jayakumari$^1$, Jochen Einbeck$^2$, John Hinde$^3$, Julien Mainguy$^4$, \\ Rafael de Andrade Moral$^1$ \\
  {\small $^1$Hamilton Institute and Department of Mathematics and Statistics, Maynooth University, Ireland} \\
  {\small $^2$Department of Mathematical Sciences, Durham University, United Kindgom} \\
  {\small $^3$School of Mathematical \& Statistical Sciences, University of Galway, Ireland} \\
  {\small $^4$Ministère de l’Environnement, de la Lutte contre les changements climatiques, de la Faune} \\
  {\small et des Parcs, Québec, Canada}}

\maketitle
 \fi

\if1\blind
{
  \bigskip
  \bigskip
  \bigskip
  \begin{center}
    {\LARGE\bf A goodness-of-fit diagnostic for count data derived 
    from half-normal plots with a simulated envelope}
\end{center}
  \medskip
} \fi

\bigskip
\begin{abstract}
Traditional methods of model diagnostics may include a plethora of graphical techniques based on residual analysis, as well as formal tests (e.g. Shapiro-Wilk test for normality and Bartlett test for homogeneity of variance). In this paper we derive a new distance metric based on the half-normal plot with a simulation envelope, a graphical model evaluation method, and investigate its properties through simulation studies. The proposed metric can help to assess the fit of a given model, and also act as a model selection criterion by being comparable across models, whether based or not on a true likelihood. More specifically, it quantitatively encompasses the model evaluation principles and removes the subjective bias when closely related models are involved. We validate the technique by means of an extensive simulation study carried out using count data, and illustrate with two case studies in ecology and fisheries research.
\end{abstract}

\noindent%
\vfill

\newpage
\spacingset{1.5} 
\section{Introduction}
\label{sec:intro}
Analyses of counts is ubiquitous to many research areas, being applied to a broad range of applied areas; for instance, the number of patients admitted to a hospital in a single day \citep{du2012use}, the progeny of insects in a biological control experiment in entomology \citep{do2013annona}, or the number of different animal species in a particular area in ecology \citep{cunningham2005modeling}. A first assumption when modelling count data typically involves the Poisson distribution~\citep{hilbe2014modeling}. This single-parameter distribution, only accounts for equidispersion, and this restrictive property may not properly accommodate a count variable's mean-variance relationship  in practical situations \citep{richards2008dealing}. It is common practice, therefore, to use extensions of the Poisson distribution that allow for more flexible mean-variance modelling, accommodating over- or under-dispersion \cite{brooks2019statistical}, as well as excess zero counts \citep{hilbe20077}.

One of the goals of fitting a model to data is to carry out inference. However, for inference about a fitted model to be reliable, goodness-of-fit tests and diagnostic analyses should be carried out to ensure that the model represents an adequate fit to the data \citep{ding2018model}. For example, residual plots can be constructed by plotting a function of the residuals, typically a scaled version of the ordinary residuals, against the predictors or fitted values \citep{tsai1998examination}. Once a model is deemed to be adequate, the inferential process typically involves hypothesis testing to compare nested models (through, e.g., likelihood-ratio tests) and assess the effects of covariates, however non-nested model selection may also be carried out by comparing measures of out-of-sample predictive performance, which include information criteria \citep{anderson2004model}.

Likelihood-based methods are ubiquitous in statistical analyses involving parametric models. However, quasi-likelihood estimation of marginal models also represents a useful approach when the inferential objective lies in modelling the mean of the data under an assumed variance-covariance structure. Examples of this approach include quasi-Poisson and quasi-binomial models where the underlying base variance function is scaled by a dispersion parameter $\phi>0$, thus accommodating over- or under-dispersion \citep{ver2007quasi,consul1990some}. This is also applicable to generalized estimating equations (GEE), which allow for the accommodation of complex dependencies in the data through prior specification of the variance-covariance structure \citep{zeger1988models}, and multivariate covariance generalized linear models (McGLM), which represent an extension to the GEE approach and allow the joint modelling of multiple responses \citep{bonat2016multivariate}. 

The marginal modelling frameworks pose challenges in terms of nested and non-nested model comparisons due to the impossibility of calculating a full likelihood measure to be used either in direct hypothesis tests, or to compute likelihood-based information criteria. This makes comparing similar, but separate, models a difficult task, and in this context graphical methods represent a complementary approach when attempting to assess  model  fit. One graphical technique that can be used to  assess empirically whether an observed data sample is a plausible realisation of a fitted model is the half-normal plot with a simulation envelope \cite{hinde1998overdispersion}. The envelope indicates the expected variability under the assumed model  and allows an assessment of the appropriateness of the model for the data at hand, indicated by the observed diagnostic half-normal plot mostly lying withing the envelope. It can be used to compare, for instance, the fit of Poisson and quasi-Poisson models, even though a full likelihood can be computed for the former model but not the latter \citep{de2017half}.

In this paper we  address these model comparison issues by introducing a goodness-of-fit metric based on distances calculated from half-normal plots with a simulated envelope. We review the graphical technique, introduce our proposed distance metric, and describe  simulation studies to explore the approach in Section \ref{sec:methods}. We present the results from the simulation studies in Section \ref{sec:results}, and show the utility of the proposed approach using two case studies in Section \ref{sec:casestudies}. Finally, we provide a discussion and conclusions in Sections \ref{sec:discussion} and \ref{sec:conclusion}, respectively.

\section{Methods}
\label{sec:methods}

\subsection{Half-normal plots with a simulated envelope}

A QQ-plot is a graphical method used in ascertaining the distribution of a sample by plotting the ordered sample quantiles versus a particular distribution's theoretical quantiles \citep{lodder1988quantile}. A normal QQ-plot is useful to identify departures from normality. A half-normal plot is similar to the normal QQ-plot, but it plots the ordered absolute values of the sample against the expected order statistics of the half-normal distribution instead of the normal; this approach is especially useful with smaller datasets. For a sample of size $n$, the expected half-normal order statistics can be approximated by $$\Phi^{-1}\left(\frac{i +n -\frac{1}{8}}{2n +\frac{1}{2}} \right), \quad i=1,\ldots,n,$$ where $\Phi^{-1}$ is the quantile function of the standard normal distribution.

\cite{atkinson1985plots} proposed the use of half-normal plots with any model-based diagnostic statistics, as well as the addition of a simulated envelope to highlight departures from what would be expected under the fitted model. The envelope is such that, if the observed data is a plausible realisation of a fitted model, for most diagnostic measures the sample values would fall within the bounds of the simulated envelope. It does not constitute a hypothesis testing procedure, however it is useful as an empirical method to detect outliers, poor goodness-of-fit, and over- or under-dispersion when analysing counts or discrete proportions \citep{hinde1998overdispersion}, depending on the particular diagnostic statistic used.

The process of constructing a $100(1-\alpha)\%$ simulated envelope is  as follows: (i) fit a model to the data and compute the ordered absolute values of a chosen diagnostic statistic (e.g., Pearson residuals, Cook's distances, or leverage values); (ii) simulate different samples from the fitted model, using the same model matrix and distribution and parameter estimates from the fit in (i) \citep{atkinson1985plots} suggests performing 19 simulations, but smoother envelopes are obtained with more simulated samples and  the \texttt{hnp} package for R software uses 99 as the default \citep{de2017half}; (iii) re-fit the model to each simulated sample and compute, for each fit, the ordered absolute values of the same diagnostic statistic used in step (i); and (iv) compute the $100\alpha/2$ and $100(1-\alpha/2)$ percentiles over the set of simulated model diagnostics for each order statistic to form the lower and upper bounds of the envelope, respectively, where typically we take $\alpha=0.05$.

\subsection{A distance measure derived from a half-normal plot with simulated envelope}

\begin{figure}[htb]
\centering
\includegraphics[width= \textwidth]{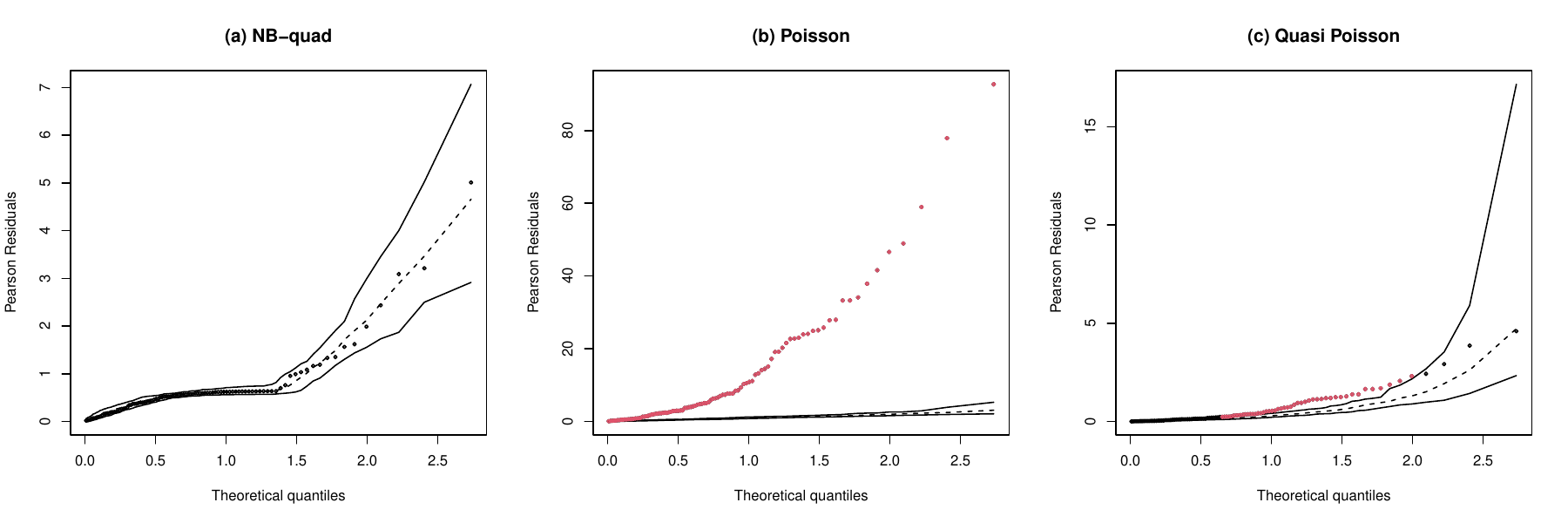}\label{fig:paperdata}
\caption{Half-normal normal plots with a simulated envelope for three different models fitted to data simulated from a negative binomial model with a quadratic variance function (NB-quad).}
\end{figure}

For illustration, Figure~\ref{fig:paperdata} displays three half-normal plots with a simulated envelope from different models fitted to data simulated from a negative binomial model with a quadratic variance function (NB-quad), i.e. from an overdispersed count model. All simulation studies and case studies discussed here use Pearson residuals for the production of the half-normal plots, however any type of diagnostic can be used in practice. The points painted in red are the ones that are falling outside the envelope. The first panel (a) shows the results from the true model, and as expected all sample residuals fall within the simulated envelope. Fitting a Poisson model (which assumes that the variance is equal to the mean), the extra variability (over-dispersion) in the data cannot be accommodated, which is evident from panel (b), with most of the sample residuals falling outside of the envelope. Panel (c) is the half-normal plot for a quasi-Poisson model (variance $\propto$ mean), and while this model can account for some of the  overdispersion, it is clear that the  linear variance function cannot fully account for the quadratic variance function of the underlying model and a considerable number of points falls outside the envelope.

Graphical goodness-of-fit assessment and model selection can be challenging in this context when closely related models are considered. Moreover, likelihood-based procedures and associated quantities are not available for marginal models such as the quasi-Poisson, although pseudo-likelihoods are sometimes used, see~\cite{bonat2016multivariate}. However, it is possible to derive an objective metric $d$ for model selection based on the sample residual values and the simulated envelope. A natural way of measuring how far the observed residuals are from the expected median behaviour is to use a quantity $d$ given by
\begin{eqnarray} \label{eq:distance}
    d=\sum_{i=1}^nd_i=\sum_{i=1}^n|r_i - m_i|^p,
\end{eqnarray}
where $r_i$ is the $i^{th}$ ordered absolute residual, $m_i$ is the median of the simulated envelope corresponding to the $i^{th}$ residual, and $p$ denotes the power used. Here, we restrict to $p\in\{1,2\}$, with $p=1$ corresponding to the absolute difference between the residual point and the median (corresponding to the $L_1$ norm), and  $p=2$ to the squared Euclidean distance.

\subsection{Simulation studies}\label{subsec:simulationstudies}

To study the properties of the proposed metric (\ref{eq:distance}), we performed simulation studies based on different underlying count distributions, including both overdispersion and zero-inflation. As we were interested in testing many distributions and the whole exercise was computationally intensive, as each simulation the calculation of $d$ requires further simulations for the formation of the simulated envelope, to speed up computation time we used a single covariate model, i.e.
\begin{eqnarray*}
    Y_i & \sim & \mathcal{D}(\mu_i,\phi,\nu) \\
    g(\mu_i) & = & \beta_0 + \beta_1x_i
\end{eqnarray*}
where $i=1,\ldots,n$ indexes the sample, $g(\cdot)$ is a link function (e.g. the log link when $\mathcal{D}$ is Poisson) for the mean $\mu_i$, $\phi$ is a dispersion parameter, and $\nu$ is a zero-inflation parameter for the parent distribution $\mathcal{D}$. We generated the values of the covariate $x_i$ using a standard normal distribution. We used three different sample sizes $n\in\{20,50,100\}$ and 6 different parent distributions $\mathcal{D}\in\{\mbox{Poisson},\mbox{Quasi-Poisson},\mbox{NB-lin},\mbox{NB-quad},\mbox{ZIP},\mbox{ZINB}\}$, where NB-lin represents a negative binomial model with a linear variance function $V(\mu_i)=\mu_i + \phi\mu_i$,  NB-quad is the negative binomial with a quadratic variance function $V(\mu_i)=\mu_i+\zeta\mu_i^2$; where $\zeta = \displaystyle\frac{1}{\phi}$, ZIP is the zero-inflated Poisson model, and ZINB is the zero-inflated negative binomial model \citep{ver2007quasi, yau2003zero, lambert1992zero}. For the Poisson and ZIP models the dispersion is  fixed, $\phi=1$; for the other models we used $\phi\in\{0.5,7\}$ to introduce scenarios of weak and strong overdispersion, respectively. For the non zero-inflated models, $\nu=0$, i.e. no zero-inflation; for the ZIP and ZINB models we used $\nu\in\{0.2,0.6\}$ to introduce weak (20\%) and strong (60\%) zero-inflation. Briefly, the zero inflation model is a mixture model that consists of a count data model and a point mass function at zero. The zero counts can arise from both the count data model and the point mass function. The count data models can be Poisson or a Negative binomial model with quadratic variance function, referred to as ZIP and ZINB, respectively. 

We performed $1,000$ simulations for each scenario. For each simulation we generated a response variable $\mathbf{y}$ from the parent distribution, and fitted all 6 models under consideration. For each model fit, we produced a half-normal plot with a simulated envelope based on Pearson residuals: $$r_i^P=\displaystyle\frac{y_i-\hat{y_i}}{\sqrt{V(\hat\mu_i)}}$$ and calculated the distance metric (\ref{eq:distance}), with $r_i \equiv r_i^P$,  using $p\in\{1,2\}$.

All computations were performed in R \citep{r2023}.  We used the \texttt{glm} function to fit the Poisson and quasi-Poisson models, and the \texttt{glm.nb} function from package MASS \citep{ripley2013package} to fit the NB-quad model, \texttt{gamlss} function of package gamlss \citep{stasinopoulos2008generalized} to fit the NB-lin model, and \texttt{zeroinfl} function of package pscl \citep{jackman2015package} to fit the ZIP and ZINB models. Half-normal plots were generated using the package hnp \citep{de2017half} with  envelopes based on 99 simulations.

All code used to produce the simulations is available at \url{https://github.com/DARSHANAJAYA/Goodness-of-fit-Distance-metric.git}.

\section{Results}\label{sec:results}

\begin{figure}[htb]
\centering
\includegraphics[width= 0.95\textwidth]{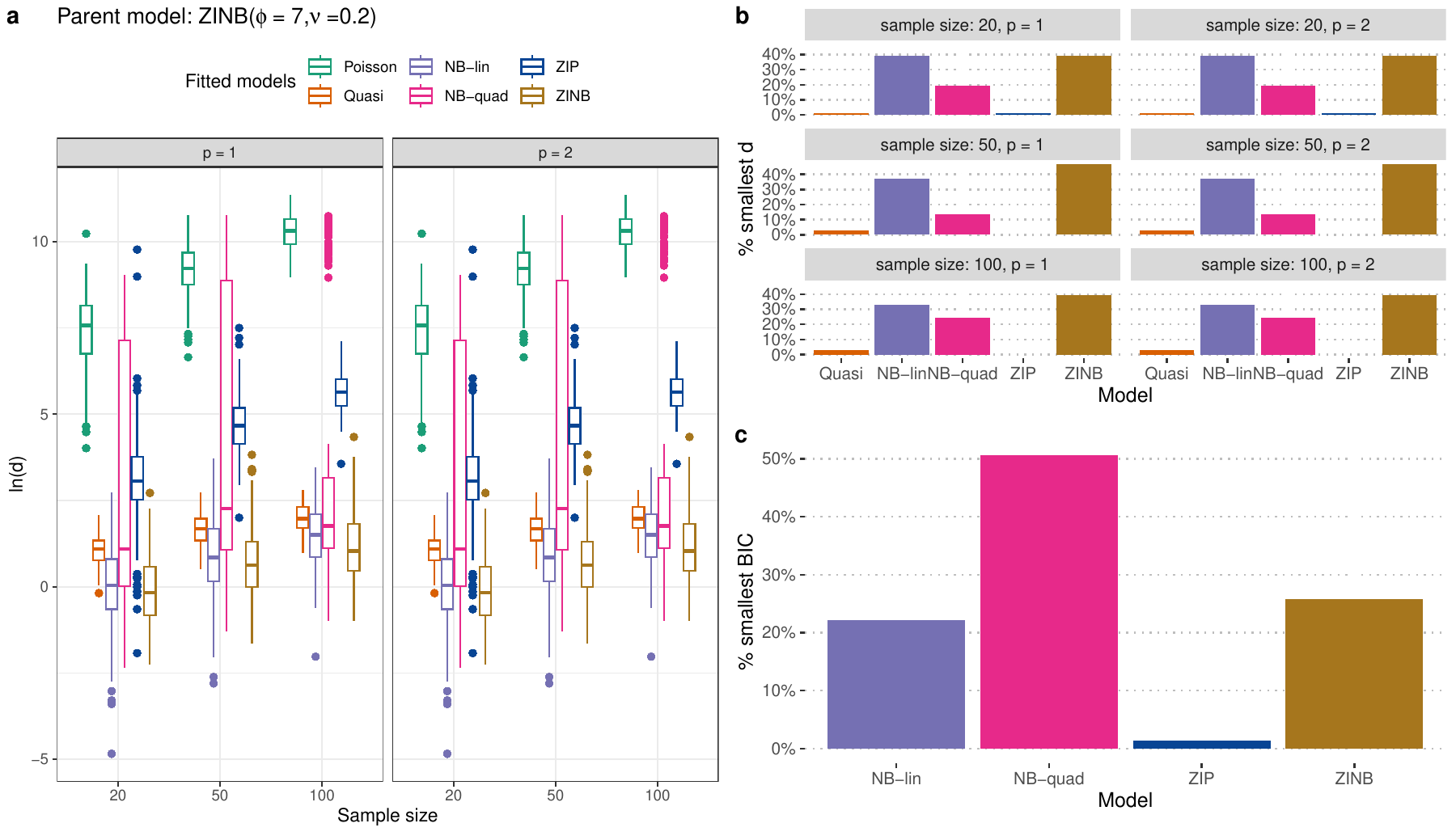}\label{fig:zinbom}
\caption{Figures generated when parent model is ZINB with a with zero inflation factor value of $0.2$ and a dispersion parameter value of $7$.Panel~(a) shows a boxplot of the base distance considered in the log scale for the fitted models. Panel~(b) shows the bar plots illustrating the number of times a particular fitted model has the distance metric computed to be the minimum in a single simulation run. Panel~(c) shows the barplot demonstrating the number of times a particular fitted model has the BIC value computed to be the minimum in a single simulation run.}
\end{figure}

Given the simulation study design, the results are divided into twelve experiments based on the different parent model considered. Results are presented in figures consisting of three plots chosen to illustrate the relevance of the proposed distance goodness-of-fit metric. Panel~(a) shows a boxplot of the proposed distance metric (\ref{eq:distance}) on the log scale for the different fitted models. Figure~\ref{fig:poisson} to Figure~\ref{fig:negbinh} include four fitted models (Poisson, quasi-Poisson, NB-lin, and NB-quad), while Figures ~\ref{fig:zipm} and  \ref{fig:zinbom} have six fitted models (Poisson, quasi-Poisson, NB-lin, NB-quad, ZIP, and ZINB). Panel~(b) shows a bar plot illustrating the number of times a particular fitted model has the distance metric computed to be the minimum in a single simulation run, i.e. points towards a particular model. Panel~(c) shows a barplot representing the number of times a particular fitted model has the smallest BIC value. Since the BIC cannot be calculated for models estimated via quasi-likelihood methods, the quasi-Poisson model is omitted from this panel. All the plots are faceted across two variables: sample size and the value of $p$. The best performing model is expected to give the minimal value for the computed distance metric in the case of panel~(a) and highest frequency level for panels~(b) and (c). For illustration purposes we have chosen to only report five experiments in this section; for the  results from the other experiments please refer to Appendix S1.

\subsection*{Simulation experiment 1}\label{simexp1}

This simulation experiment assessed the effectiveness of the distance metric when the parent model is an equidispersion one (Poisson). The inherent randomness of the simulation causes the dataset to be not exactly equidispersed, but either slightly overdispersed or underdispersed. While Figure~\ref{fig:poisson}(a) and Figure~\ref{fig:poisson}(b) show that the quasi-Poisson model performs better than the competitor models with the lowest distance value, although the BIC favours the (true) Poisson model (Figure~\ref{fig:poisson}(c)) in experiment 1.

\begin{figure}[htb]
\centering
\includegraphics[width= 0.95\textwidth]{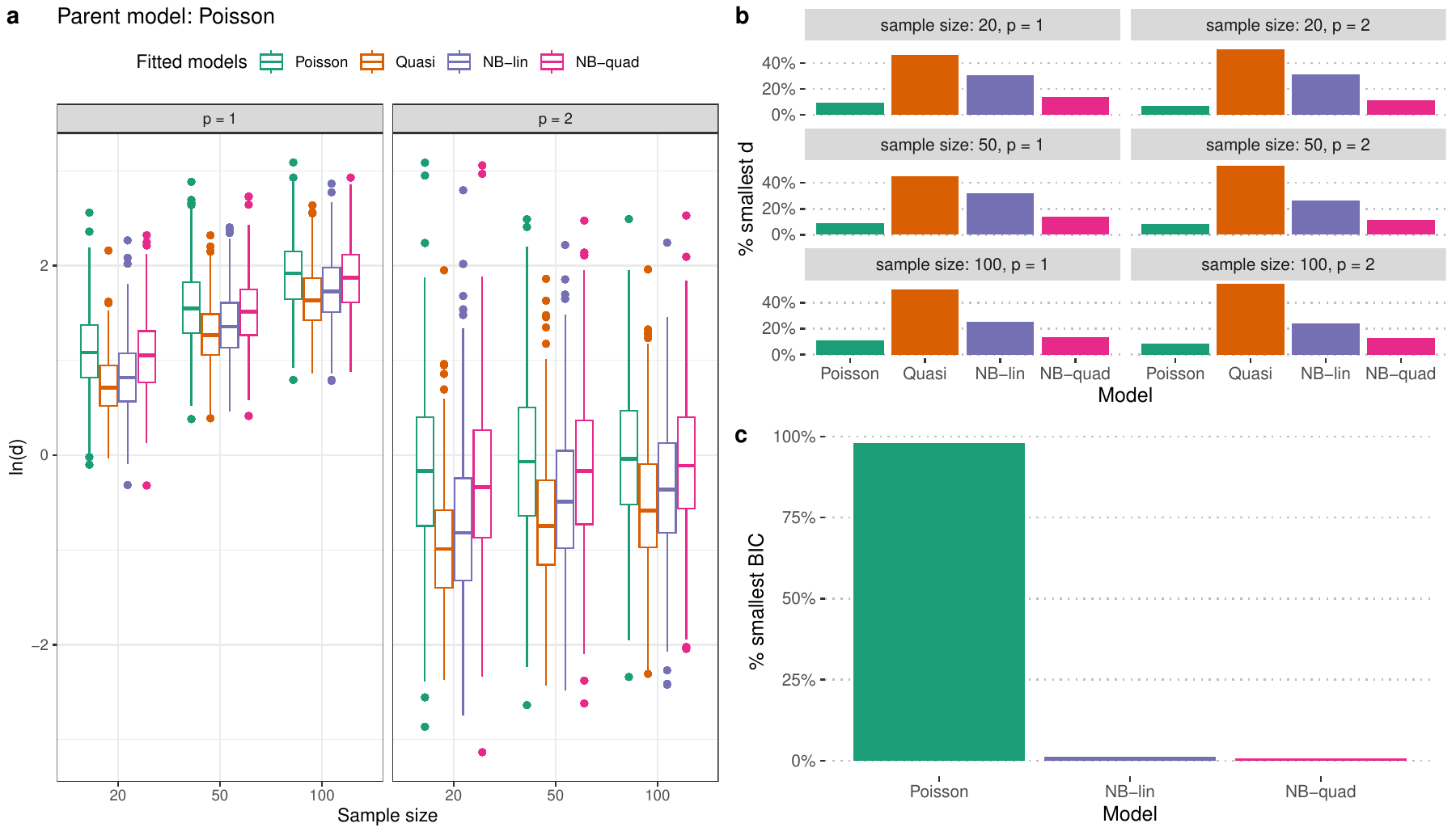}\label{fig:poisson}
\caption{Figures generated when the parent model is the Poisson.}
\label{fig:poisson}
\end{figure}

\subsection*{Simulation experiment 2}\label{simexp2}

\begin{figure}[htb]
\centering
\includegraphics[width= 0.95\textwidth]{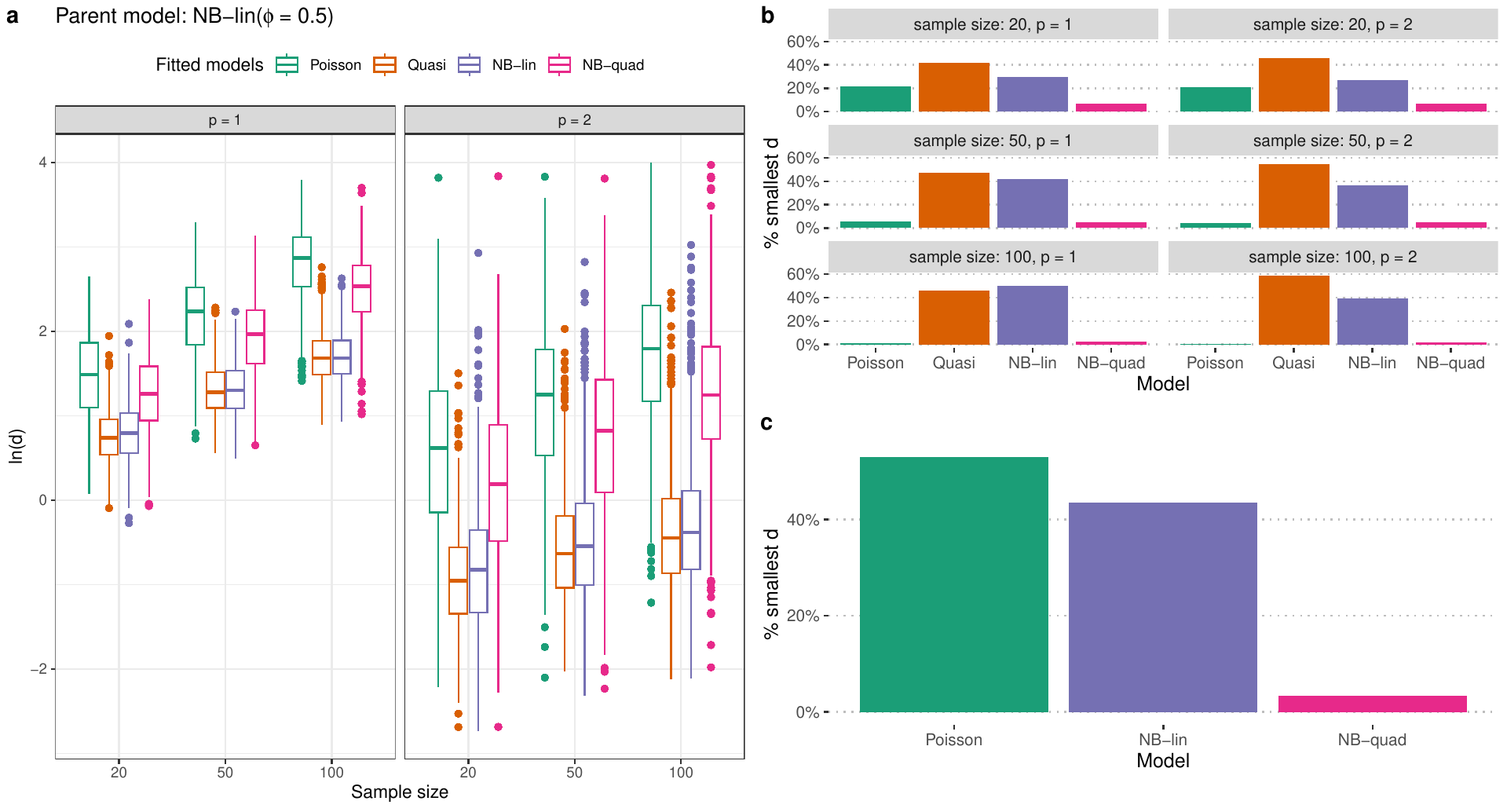}
\caption{Figures generated when the parent model is the NB-lin with a dispersion parameter value of $0.5$.}
\label{fig:negbinmlin}
\end{figure}

This experiment portrays the results when the parent model under consideration is one that can accommodate overdispersion, a negative binomial model with a linear variance function (NB-lin). The dispersion parameter chosen was $0.5$ and is deemed a case of mild overdispersion. The results show that the quasi-Poisson and NB-lin perform the best, as is expected since both of these models have a linear variance function. Figure~\ref{fig:negbinmlin}(a) and (b) show that using the absolute differences (equivalent to the $L_1$ norm) improves the performance of the distance metric in identifying the true parent model. In Figure~\ref{fig:negbinmlin}(c) BIC favours the Poisson model as the best-fitting one.

\subsection*{Simulation experiment 3}\label{simexp3}

\begin{figure}[htb]
\centering
\includegraphics[width= 0.95\textwidth]{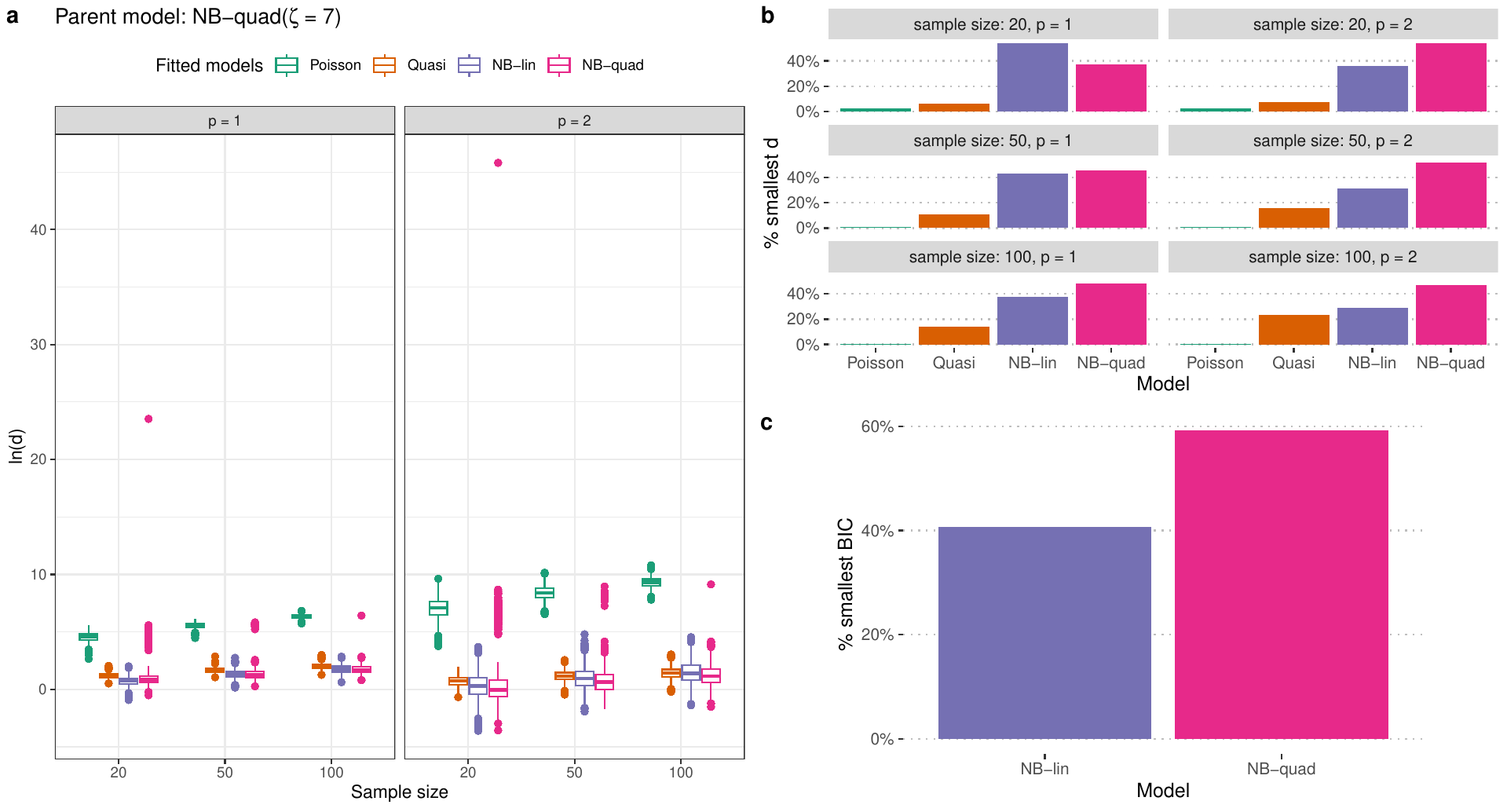}
\caption{Figures generated when the parent model is the NB-quad with a dispersion parameter value of $7$.}
\label{fig:negbinh}
\end{figure}

This experiment illustrates the results when the parent model is a GLM with a quadratic variance function (NB-quad), a model that  can accommodate overdispersion. In this specific case, the dispersion parameter value is chosen to be $7$ and is considered an instance of strong overdispersion. Figure~\ref{fig:negbinh}(a) and (b) show that the proposed goodness-of-fit diagnostic picks the best model as the parent model at  sample sizes of $50$ and $100$, which is in agreement with the BIC (Figure~\ref{fig:negbinh}(c)).

\subsection*{Simulation experiment 4}\label{simexp4}

\begin{figure}[htb]
\centering
\includegraphics[width= 0.95\textwidth]{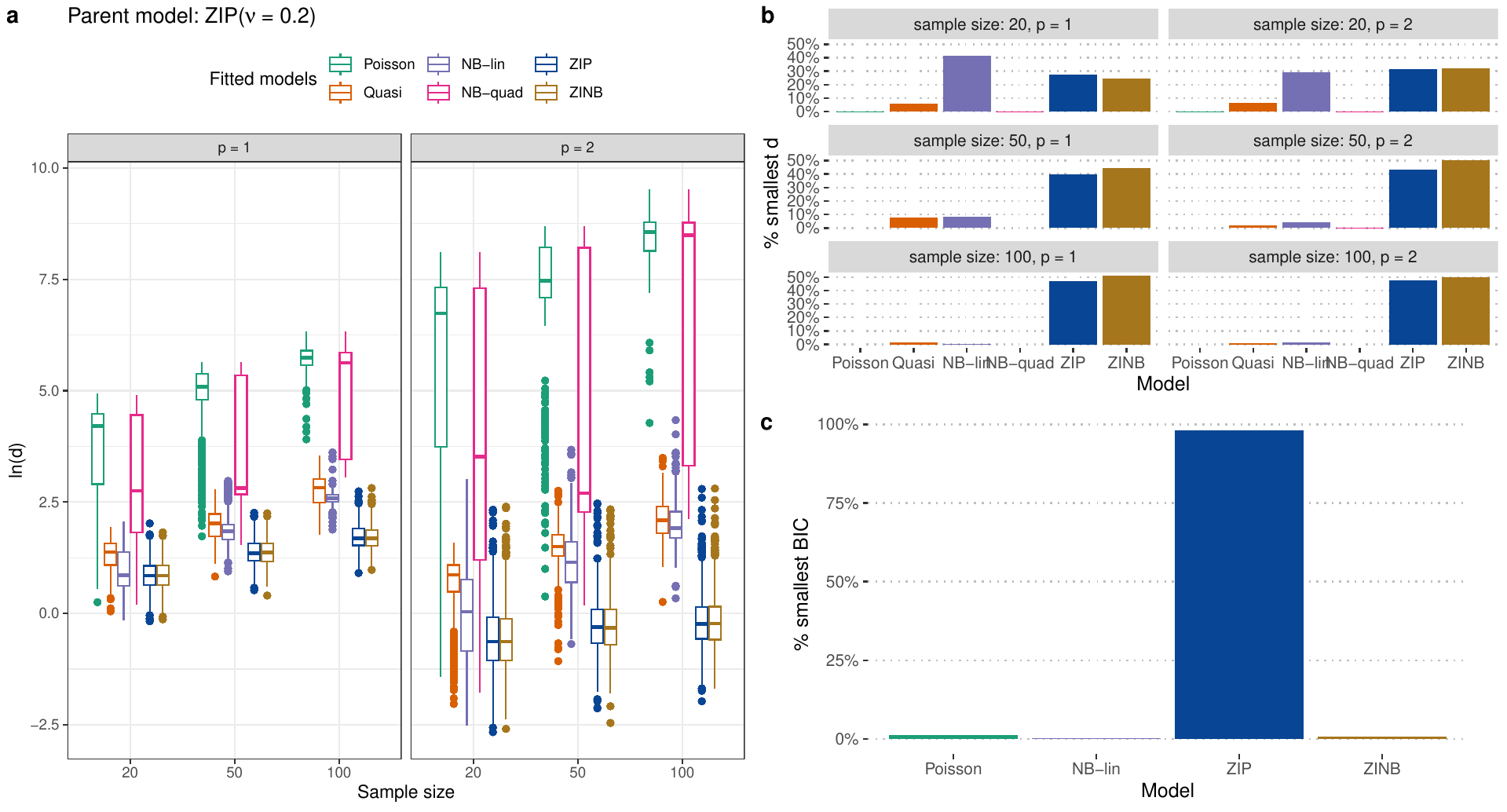}
\caption{Figures generated when the parent model is the ZIP with  a zero inflation factor value of $0.2$.}
\label{fig:zipm}
\end{figure}

Simulation experiment 4 involves a ZIP model as the parent model with a zero-inflation parameter value of $0.2$,  a low level of zero-inflation. The results  in Figure~\ref{fig:zipm}(a) and (b) show that the distance metric selects both the ZIP and ZINB models as the best performing models. This likely occurs due to the randomness in the simulated datasets where an inherent overdispersion can occur by chance in the simulation. In Figure~\ref{fig:zipm}(c), BIC favours the parent model, rejecting the additional complexity of the ZINB model.

\subsection*{Simulation experiment 5}\label{simexp5}

\begin{figure}[htb]
\centering
\includegraphics[width= \textwidth]{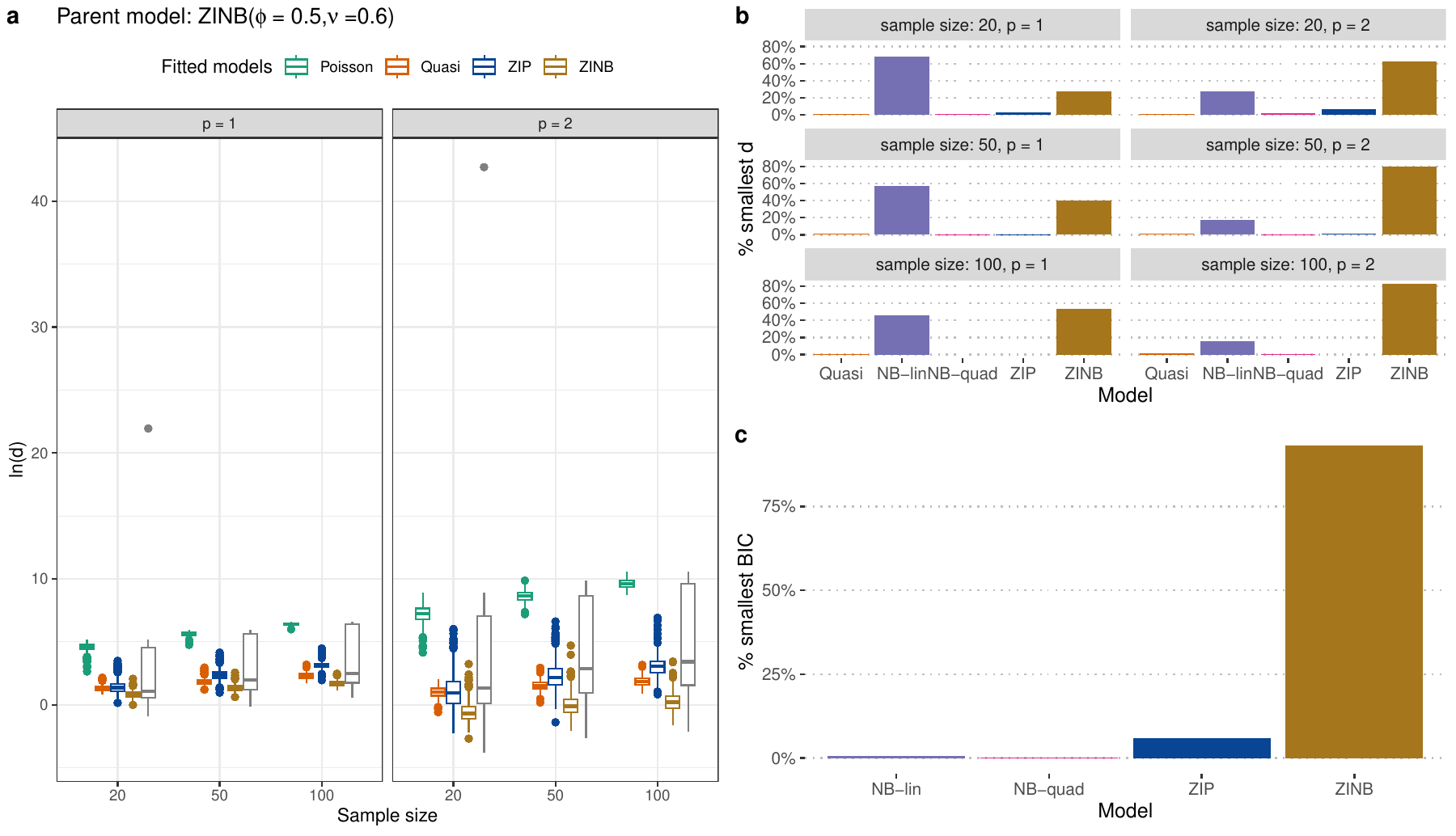}
\caption{Figures generated when the parent model is ZINB with a with a zero inflation factor value of $0.6$ and a dispersion parameter value of $0.5$.}
\label{fig:zinbuh}
\end{figure}

This experiment uses a ZINB model as the parent model with a dispersion parameter value of $0.5$, and zero inflation parameter value of $0.6$, which provides a scenario where the parent model is zero-inflated negative binomial with mild overdispersion and high zero-inflation. Figures~\ref{fig:zinbuh}(a) and (b) show that the distance metric shows better performance in selecting the true model when the squared Euclidean distance is considered. In Figure~\ref{fig:zinbuh}(c), BIC favours the parent model as the true model in all instances.

\section{Case Studies}\label{sec:casestudies}

We now present the analysis of two case studies using the distance metric proposed in equation~(\ref{eq:distance}) to aid goodness-of-fit assessment and model selection. All R code and data are available at \url{https://github.com/DARSHANAJAYA/Goodness-of-fit-Distance-metric.git}.

\subsection{Spider data}\label{subsec:spider}

\begin{figure}[htb]
\centering
\includegraphics[width= \textwidth]{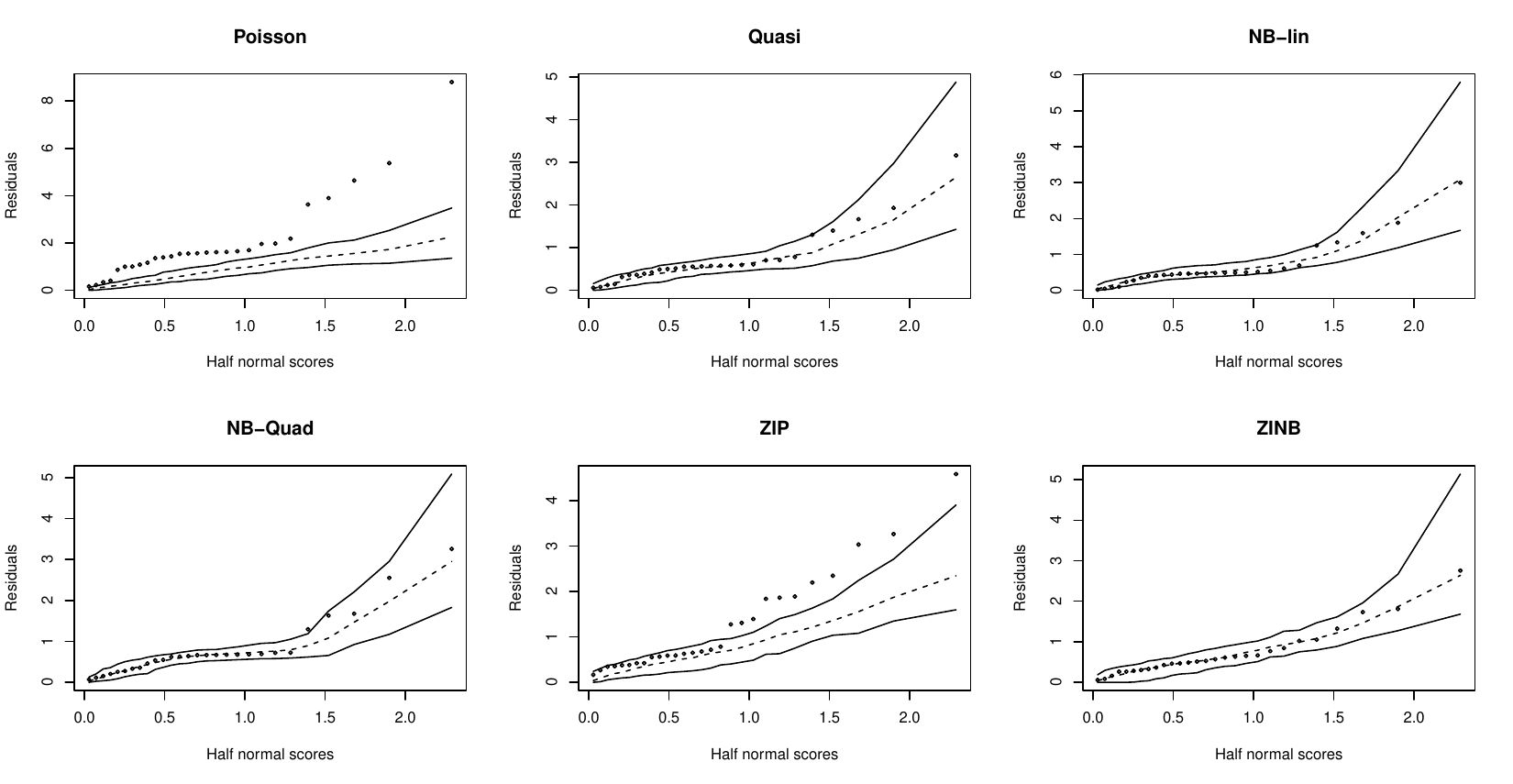}
\caption{Spider data: half-normal plots with a simulated envelope for the Poisson, quasi-Poisson, NB-lin, NB-quad, ZIP and ZINB models.}
\label{fig:hnpspider}
\end{figure}

This dataset is from the paper by \citep{smeenk1974correlations} and is also included in the R package mvabund \citep{wang2012mvabund}. We considered the count of the hunting spiders from the species \textit{Alopecosa accentuata} that were caught in 100 pitfall traps (1-m radius) over a period of 60 weeks, taking the soil dry mass as a covariate. The soil dry mass was $\log(x +1)$ transformed prior to model fitting. We fitted the Poisson, quasi-Poisson, NB-lin, NB-quad, ZIP, and ZINB models. The half-normal plots corresponding to the fitted models were constructed $100$ times, of which one is presented for each model in Figure~\ref{fig:hnpspider}. The process of producing the half-normal plot $100$ times is not necessary when employing this technique, however we do so here to understand the degree of uncertainty of the distance metric in this case study. The median of the distance metric was estimated from the $100$ iterations and was used to assess the fit of the models considered (Table~\ref{table:spiderdistances}). The distance metric favours NB-Lin for the squared Euclidean distance and ZINB for the $L_1$ 
 norm and this is explained as the estimated  zero inflation parameter of the ZINB model is $2.75 \times 10^{-5}$, which can be considered meaningless and so close to no zero-inflation. The principle of parsimony would favour NB-lin, and thus the  proposed distance metric shows that NB-lin captures the patterns in the data better than the other models considered, which is in line with the BIC values. The estimates of the soil mass from the summary of the NB-lin indicated that there is a higher number of hunting spiders associated with a lower level of soil dry mass. 

\begin{table}[]
\begin{tabular}{llllllll}
\hline
\multirow{2}{*}{} & \multicolumn{3}{c}{\textbf{p = 1}} & \multicolumn{3}{c}{\textbf{p = 2}} & \multicolumn{1}{c}{\multirow{2}{*}{\textbf{BIC}}} \\ \cline{2-7}
 & \multicolumn{1}{l}{\textbf{Median}} & \multicolumn{1}{l}{\textbf{IQR}} & \textbf{SD} & \multicolumn{1}{l}{\textbf{Median}} & \multicolumn{1}{l}{\textbf{IQR}} & \textbf{SD} & \multicolumn{1}{c}{} \\ \hline
\textbf{Poisson} & \multicolumn{1}{l}{34.53} & \multicolumn{1}{l}{0.42} & 0.34 & \multicolumn{1}{l}{91.06} & \multicolumn{1}{l}{2.01} & 1.52 & 246.18 \\ 
\textbf{Quasi Poisson} & \multicolumn{1}{l}{2.66} & \multicolumn{1}{l}{0.29} & 0.22 & \multicolumn{1}{l}{0.71} & \multicolumn{1}{l}{0.19} & 0.16 & \multicolumn{1}{c}{-} \\ 
\textbf{NB - lin} & \multicolumn{1}{l}{2.20} & \multicolumn{1}{l}{0.17} & 0.12 & \multicolumn{1}{l}{0.33} & \multicolumn{1}{l}{.04} & 0.05 & \multicolumn{1}{c}{141.83} \\ 
\textbf{NB - quad} & \multicolumn{1}{l}{2.95} & \multicolumn{1}{l}{0.17} & 0.15 & \multicolumn{1}{l}{1.13} & \multicolumn{1}{l}{0.14} & 0.11 & 148.82 \\ 
\textbf{ZIP} & \multicolumn{1}{l}{13.16} & \multicolumn{1}{l}{0.34} & 0.28 & \multicolumn{1}{l}{14.21} & \multicolumn{1}{l}{0.64} & 0.43 & 194.09 \\ 
\textbf{ZINB} & \multicolumn{1}{l}{1.29} & \multicolumn{1}{l}{0.16} & 0.11 & \multicolumn{1}{l}{0.14} & \multicolumn{1}{l}{0.04} & 0.03 & 141.83 \\ \hline
\end{tabular}

\caption{Median, interquartile range(IQR) and standard deviation (SD) of the distance metric (eq. \ref{eq:distance}) calculated with $p=1$ and $p=2$, obtained from 100 half-normal plots generated for six different models fitted to the spider data, and associated BIC values.} \label{table:spiderdistances}
\end{table}

\subsection{Walleye data}\label{subsec:walleye}

\begin{figure}[htb]
\centering
\includegraphics[width= \textwidth]{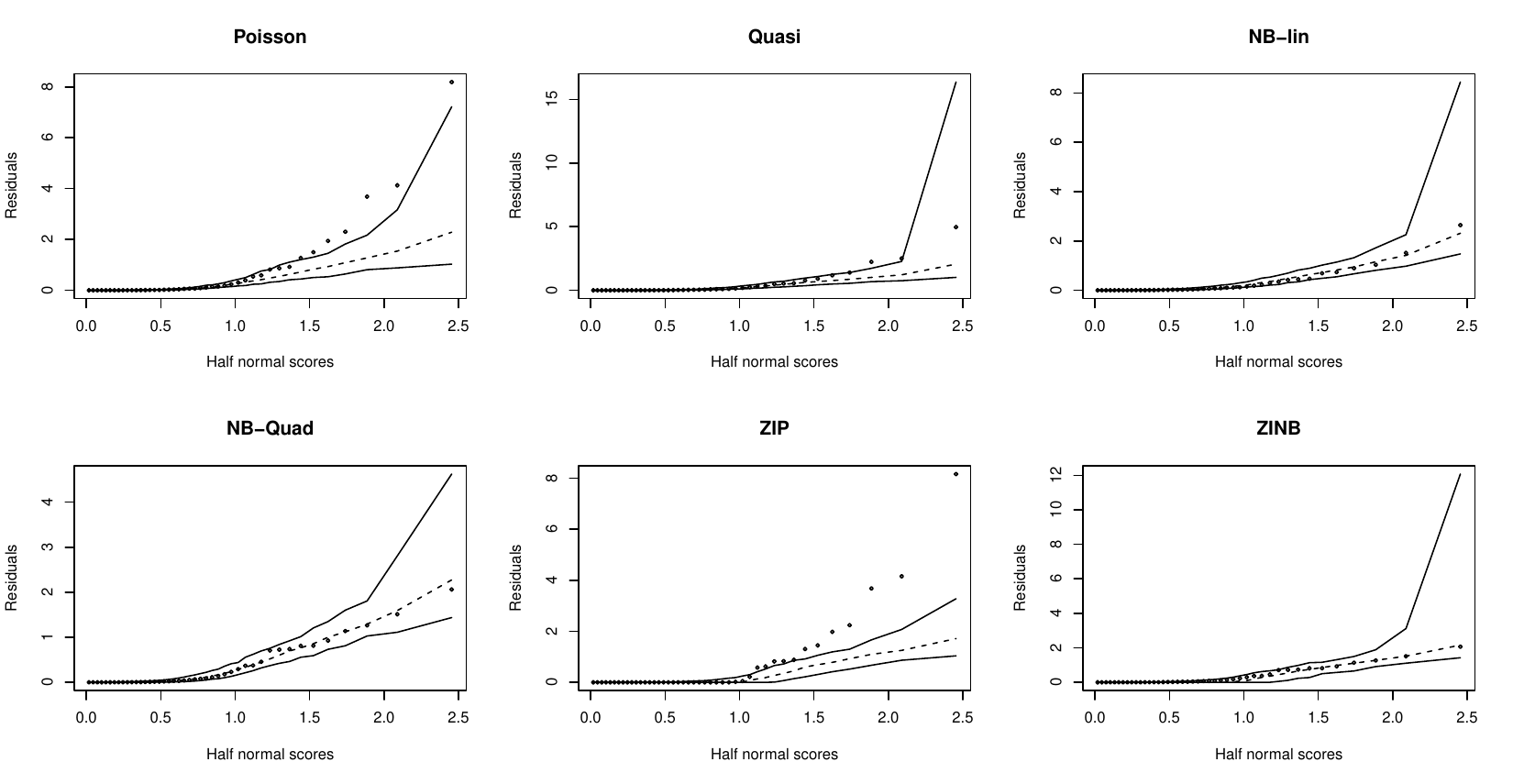}
\caption{Walleye data: half-normal plots with a simulated envelope for the Poisson, quasi-Poisson, NB-lin, NB-quad, ZIP and ZINB models.}
\label{fig:hnpfisheries}
\end{figure}

In this  case study we used a subset of the catch curve data analysed by \citep{mainguy2021improved}. The catch curve data is age frequency data that is used to estimate the instantaneous mortality of fish. We looked at the walleye (\textit{Sander vitreus}) gillnet survey data from the year 2012 from the Baskatong reservoir, Qu\'ebec, Canada. The response variable considered is the count of fish which is modelled according to age fitted as a predictor variable, such that the rate at which counts decrease with age can be used to estimate mortality. Here we fitted the same six models as for the analysis of the spider data and produced 100 half-normal plots with a simulated envelope for each model fit (one is displayed for each model in Figure~\ref{fig:hnpfisheries}). The median distance metric is shown in Table~\ref{table:fisheriesdistances}. Using the squared Euclidean distance, the distance metric favours the NB-quad model, whereas for the $L_1$ norm it favours the ZINB model, with the NB-quad ranked closely. This makes sense since if the NB-quad provides a good fit, then the ZINB also should even in the case zero-inflation is not present. For the walleye data, the zero-inflation parameter is estimated as $1.42 \times 10^{-6}$, which reflects  very low or no  zero-inflation. The BIC favours the NB quad model. This is in line with the findings from \citep{mainguy2021improved}, indicating the extra variability in the walleye data was best accommodated by the NB-quad model. The proposed distance metric has an added advantage of avoiding subjective bias that is present in the graphical model selection method apparent from this case study.

\begin{table}[]
\begin{tabular}{llllllll}
\hline
\multirow{2}{*}{} & \multicolumn{3}{c}{\textbf{p = 1}} & \multicolumn{3}{c}{\textbf{p = 2}} & \multicolumn{1}{c}{\multirow{2}{*}{\textbf{BIC}}} \\ \cline{2-7}
 & \multicolumn{1}{l}{\textbf{Median}} & \multicolumn{1}{l}{\textbf{IQR}} & \textbf{SD} & \multicolumn{1}{l}{\textbf{Median}} & \multicolumn{1}{l}{\textbf{IQR}} & \textbf{SD} & \multicolumn{1}{c}{} \\ \hline
\textbf{Poisson} & \multicolumn{1}{l}{16.32} & \multicolumn{1}{l}{0.37} & 0.28 & \multicolumn{1}{l}{53.90} & \multicolumn{1}{l}{2.25} & 1.65 & 164.76 \\ 
\textbf{Quasi Poisson} & \multicolumn{1}{l}{7.12} & \multicolumn{1}{l}{0.36} & 0.25 & \multicolumn{1}{l}{12.02} & \multicolumn{1}{l}{0.98} & 0.77 & \multicolumn{1}{c}{-} \\ 
\textbf{NB - lin} & \multicolumn{1}{l}{1.21} & \multicolumn{1}{l}{0.25} & 0.19 & \multicolumn{1}{l}{0.13} & \multicolumn{1}{l}{0.08} & 0.07 & \multicolumn{1}{c}{104.99} \\ 
\textbf{NB - quad} & \multicolumn{1}{l}{1.098} & \multicolumn{1}{l}{0.17} & 0.13 & \multicolumn{1}{l}{0.13} & \multicolumn{1}{l}{0.06} & 0.05 & 93.78 \\ 
\textbf{ZIP} & \multicolumn{1}{l}{18.04} & \multicolumn{1}{l}{0.35} & 0.27 & \multicolumn{1}{l}{60.36} & \multicolumn{1}{l}{1.43} & 1.09 & 171.53 \\ 
\textbf{ZINB} & \multicolumn{1}{l}{2.57} & \multicolumn{1}{l}{0.31} & 0.23 & \multicolumn{1}{l}{0.35} & \multicolumn{1}{l}{0.09} & 0.07 & 101.35 \\ \hline
\end{tabular}
\caption{Median, interquartile range(IQR) and standard deviation (SD) of the distance metric (eq. \ref{eq:distance}) calculated with $p=1$ and $p=2$, obtained from 100 half-normal plots generated for six different models fitted to the walleye data, and associated BIC values.} \label{table:fisheriesdistances}

\end{table}
\section{Discussion}\label{sec:discussion}

This paper focussed on defining a quantitative summary to the qualitative graphical model selection and goodness-of-fit assessment method known as a half-normal plot with a simulated envelope. A simple and effective distance metric was introduced that could capture how far the observed data deviates from the expected behaviour according to the fitted model. We considered  two forms of distances through the power coefficient $p$ and found that they  were useful in differentiating the fit of count data models with a linear variance function (quasi-Poisson or NB-lin) and those with a quadratic variance function (NB-quad, ZINB). We carried out further simulation studies to understand the effectiveness of adding a measure of envelope width in the distance (\ref{eq:distance}), as well as an extra penalty term  when a residual falls outside the envelope. However, these seemed to have no real impact on the final conclusions, and therefore were not used in our formulation. See the appendix for the alternative formulation and results from the additional simulation studies.

When overdispersed counts are analysed, which corresponds to a commonly-encountered situation in ecology \citep{richards2008dealing}, determining whether such extra variation should be modelled as a linear or quadratic function of the mean is not trivial \citep{ver2007quasi}. When only applying a correction factor to the standard error through a quasi-Poisson approach \citep{knape2016decomposing}, or quasi-binomial one when modelling overdispersed discrete proportions instead \citep{bolnick2014major}, this may sometimes be sufficient to account for the detected overdispersion and then use the quasi-AIC to identify the best-fitting model. However, using a scaling parameter to directly model how data are dispersed, such as the one used with the NB-quad and NB-lin, may offer a better approach than relying on the former \citep{shadish2014using,harrison2015comparison}. With the proposed distanced-based method described in this paper, identifying which of overdispersed model extensions can now be assessed on a similar basis from not only an adequacy (i.e., goodness-of-fit) perspective, but also to help with model selection to identify the one that should be retained. As such, the proposed metric which however only assess the fit without accounting for model complexity, can then be complemented with other commonly-used ones, such an information criteria like the BIC that was used in this study as a complement to further fine-tune the model selection process.

Future work arising from this study would be to explore the impact of a mis-specified link function and  missing covariates  on the methodology and the behaviour of the proposed distance metric. Another scenario that could be be further investigated  are the response patterns when mixed models are included in the study.

\section{Conclusion} \label{sec:conclusion}
The proposed distance metric framework provides a competitive  goodness-of-fit diagnostic to check the adequacy  of count data models. This was validated by a comprehensive simulation study that showed that our proposed metric is comparable, or in some cases superior, to BIC in identifying the true model that generated the data. It represents, therefore, a complementary approach in goodness-of-fit assessment that adds an objective measure of fit when comparing half-normal plots with a simulated envelope from competing models, especially when results are similar and is particularly useful when likelihood-based methods are not available.

\section*{Acknowledgments}

This publication has emanated from research supported by a grant from Science Foundation Ireland, under Grant number 18/CRT/6049.

\bibliographystyle{agsm}

\bibliography{Bibliography-MM-MC}

\newpage

\section*{Appendix}\label{sec:Appendix}

\subsection*{Appendix S1}

This appendix displays the figures related to the simulation scenarios not presented in the main body of text of the paper.

\begin{figure}[htb]
\centering
\includegraphics[width= 0.95\textwidth]{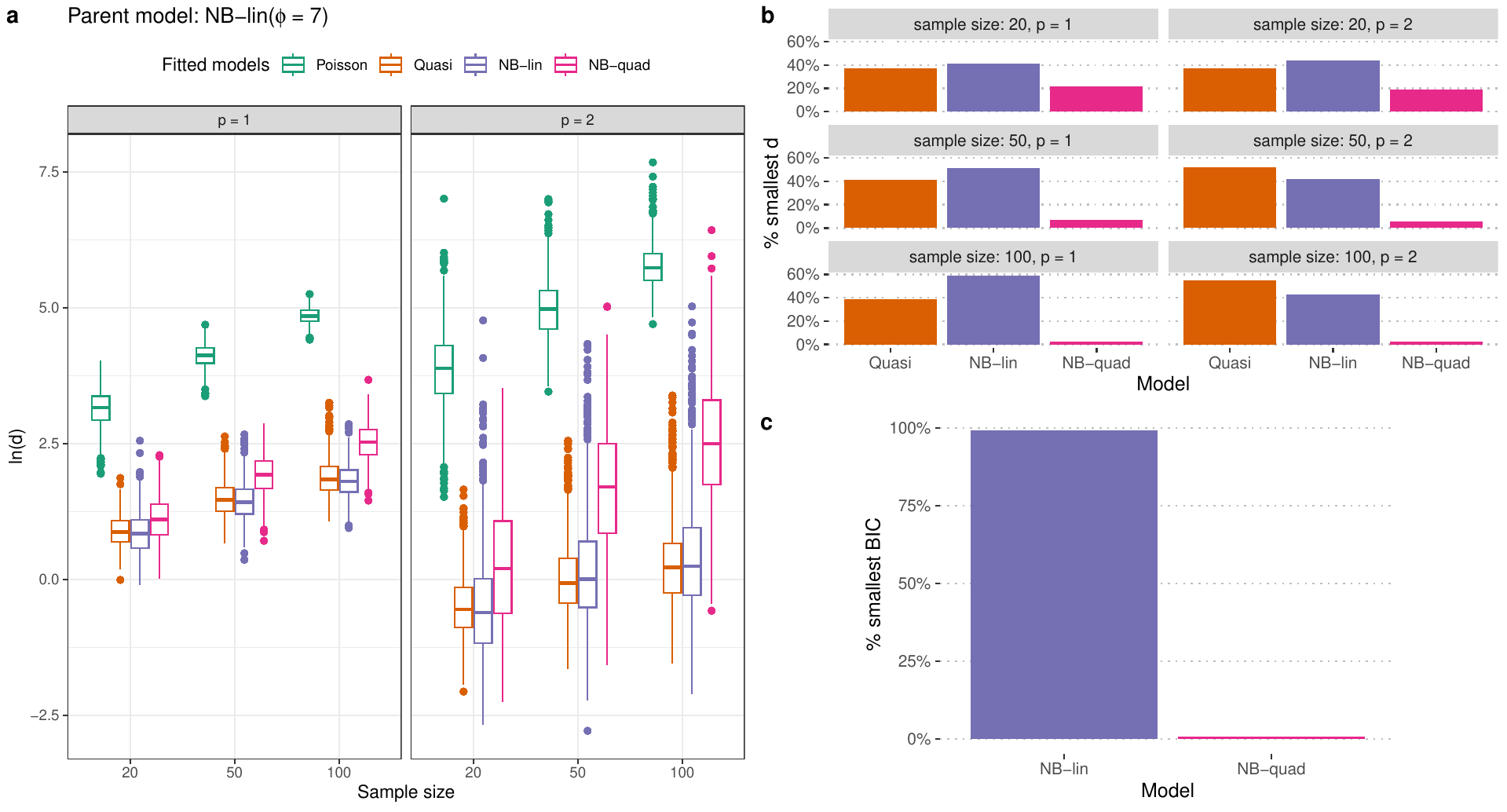}\label{fig:negbinhlin}
\caption{Figures generated when parent model is NB-lin with a with an dispersion parameter value of $7$. Panel~(a) shows a boxplot of the base distance considered in the log scale for the fitted model. Panel~(b) shows the bar plot illustrating the number of times a particular fitted model has the distance metric computed to be the minimum in a single simulation. Panel~(c) shows the barplot demonstrating the number of times a particular fitted model has the BIC value computed to be the minimum in a single simulation.}
\end{figure}

\begin{figure}[htb]
\centering
\includegraphics[width= 0.95\textwidth]{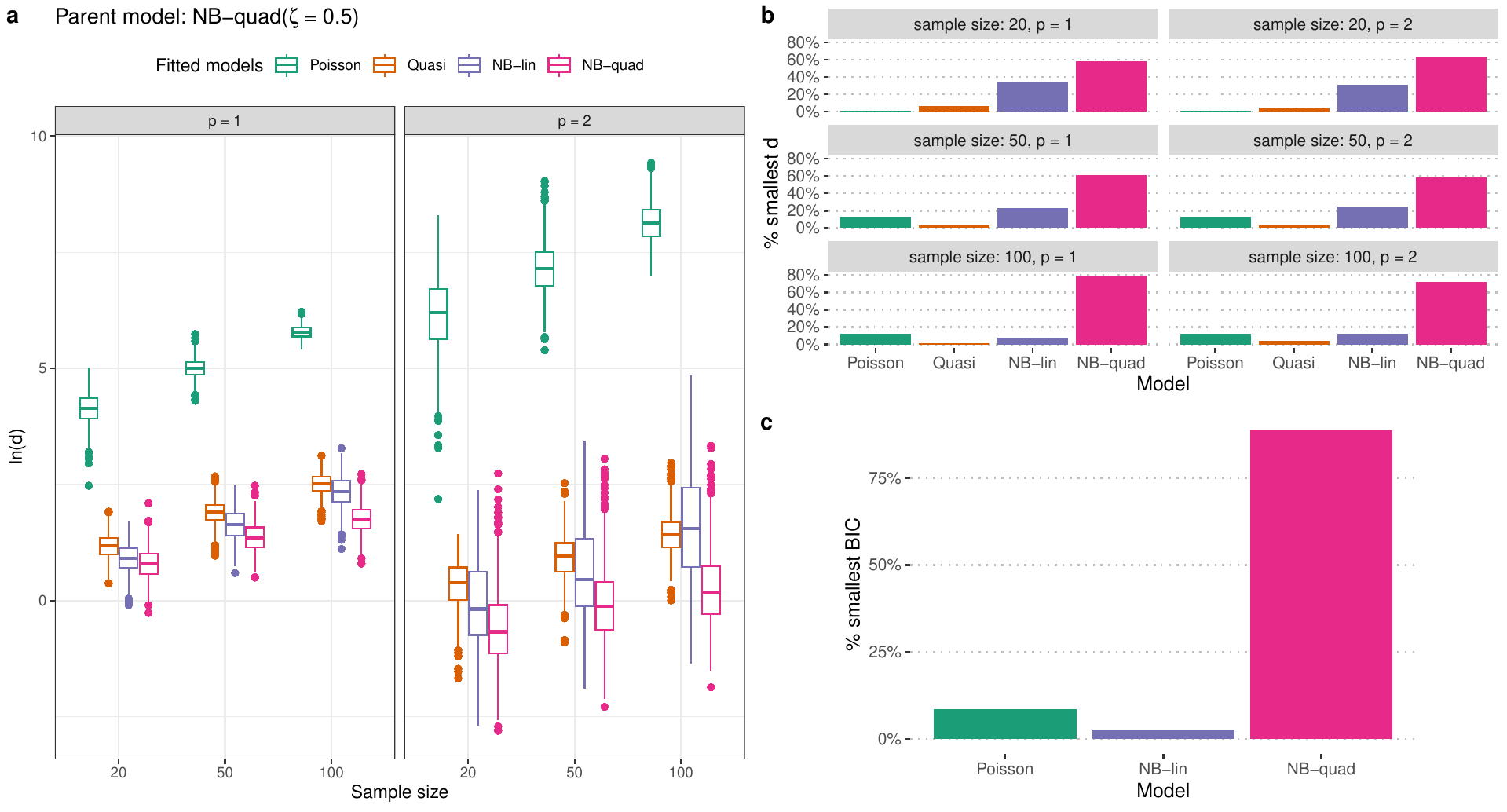}\label{fig:negbinm}
\caption{Figures generated when the parent model is NB-quad with a dispersion parameter value of $0.5$. Panel~(a) shows a boxplot of the base distance considered in the log scale for the fitted models. Panel~(b) shows the bar plots illustrating the number of times a particular fitted model has the distance metric computed to be the minimum in a single simulation run. Panel~(c) shows the barplot demonstrating the number of times a particular fitted model has the BIC value computed to be the minimum in a single simulation run.}
\end{figure}

\begin{figure}[htb]
\centering
\includegraphics[width= 0.95\textwidth]{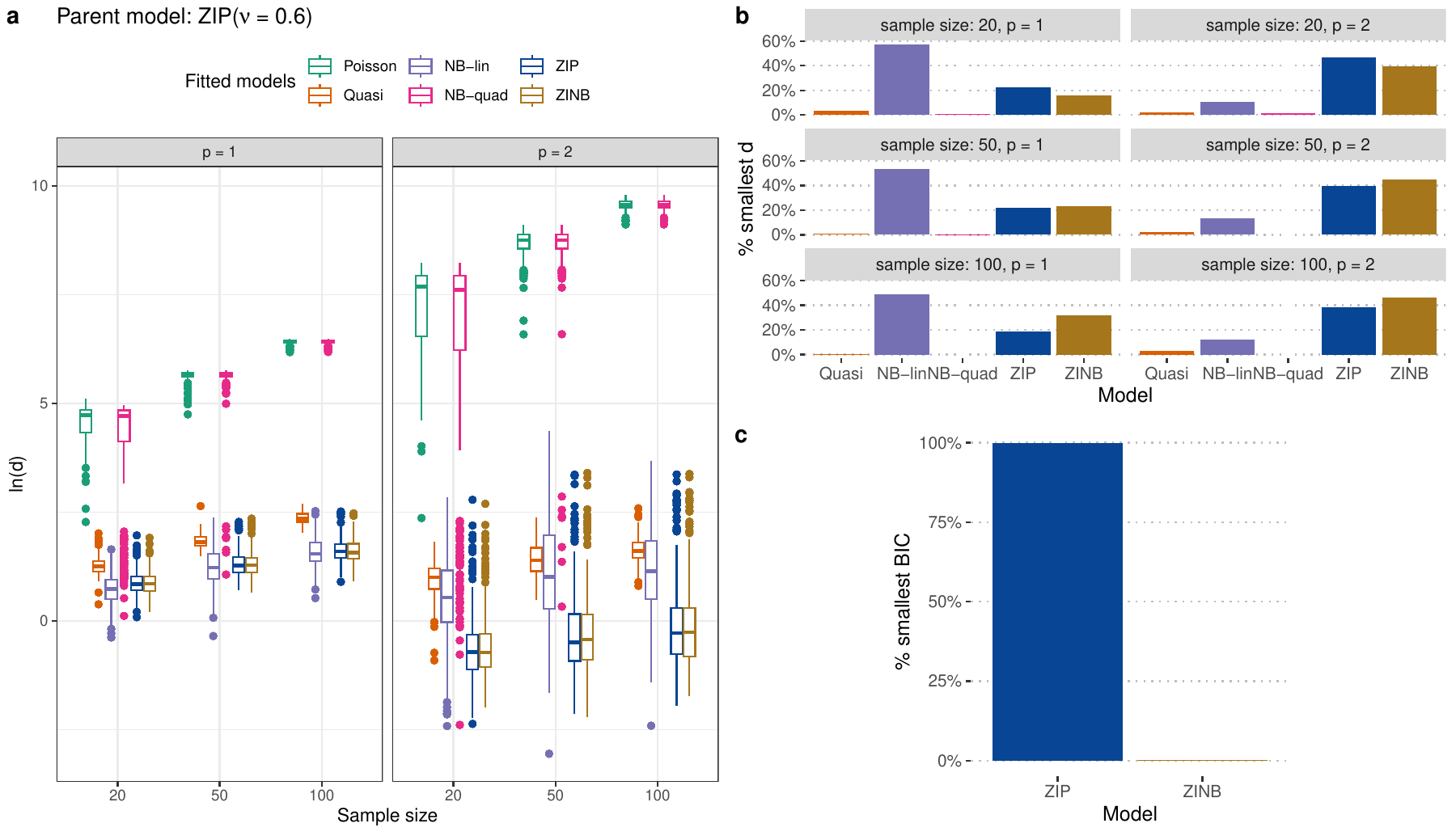}\label{fig:ziph}
\caption{Figures generated when parent model is ZIP with a with zero inflation factor value of $0.6$. Panel~(a) shows a boxplot of the base distance considered in the log scale for the fitted models. Panel~(b) shows the bar plots illustrating the number of times a particular fitted model has the distance metric computed to be the minimum in a single simulation run. Panel~(c) shows the barplot demonstrating the number of times a particular fitted model has the BIC value computed to be the minimum in a single simulation run.}
\end{figure}

\begin{figure}[htb]
\centering
\includegraphics[width= 0.95\textwidth]{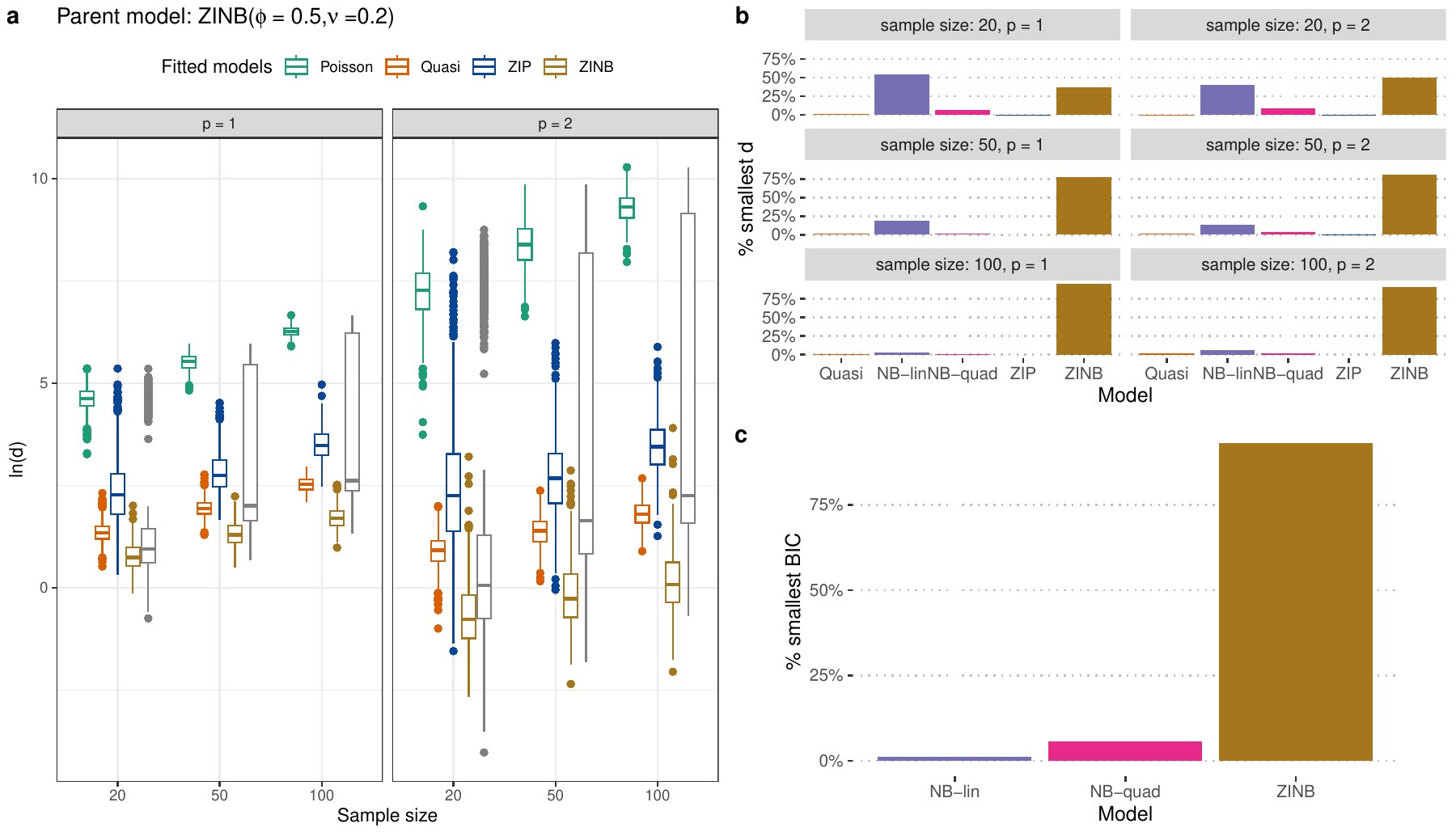}\label{fig:zinbum}
\caption{Figures generated when parent model is ZINB with a with zero inflation factor value of $0.2$ and a dispersion parameter value of $0.5$. Panel~(a) shows a boxplot of the base distance considered in the log scale for the fitted models. Panel~(b) shows the bar plots illustrating the number of times a particular fitted model has the distance metric computed to be the minimum in a single simulation run. Panel~(c) shows the barplot demonstrating the number of times a particular fitted model has the BIC value computed to be the minimum in a single simulation run.}
\end{figure}

\begin{figure}[htb]
\centering
\includegraphics[width= 0.95\textwidth]{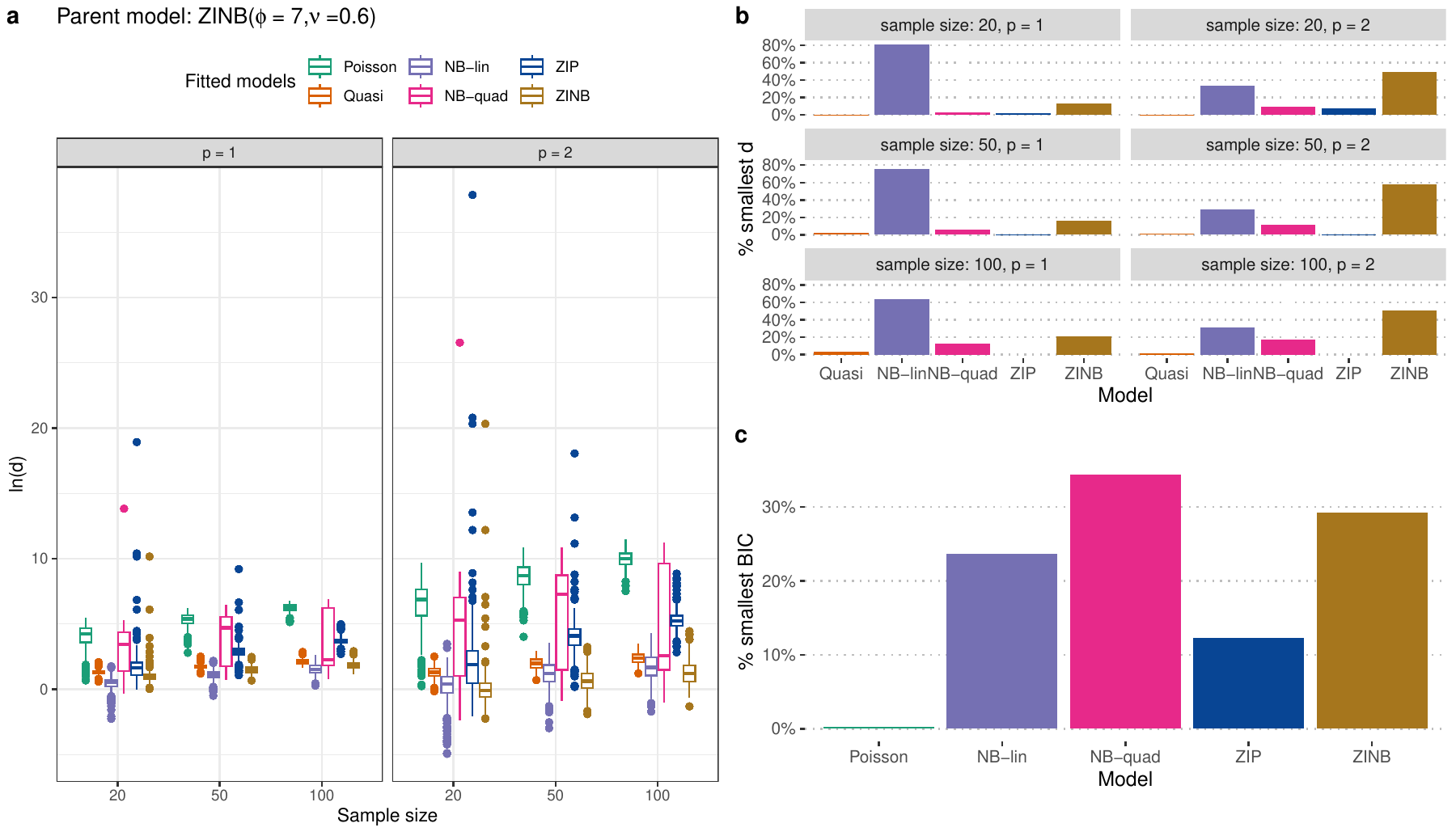}\label{fig:zinboh}
\caption{Figures generated when parent model is ZINB with a with zero inflation factor value of $0.6$ and a dispersion parameter value of $7$. Panel~(a) shows a boxplot of the base distance considered in the log scale for the fitted models. Panel~(b) shows the bar plots illustrating the number of times a particular fitted model has the distance metric computed to be the minimum in a single simulation run. Panel~(c) shows the barplot demonstrating the number of times a particular fitted model has the BIC value computed to be the minimum in a single simulation run.}
\end{figure}

The under dispersed data was generated by forcing the Poisson distribution to halve  the values produced by rpois function from the stats package to induce the deisred mean-variance relationship. Among the models considered quasi- Poisson model realises the underdispersed data and the distance metric also shows the same. 

\newpage

\subsection*{Appendix S2}\label{appendix:S2}

This appendix discusses an extension to the simulation study with two additional actors integrated into the distance metric proposed in \ref{subsec:simulationstudies}. 

We constructed a statistic based on distances from the residual points to parts of the envelope
$\mathcal{E}_i=\{x\in\mathcal{R}|x\in(l_i,u_i)\}$, namely the envelope median $m_i$, upper ($u_i$) and lower ($l_i$) bounds. The statistic is given by: $$d=\sum_{i=1}^nd_i=\mathlarger{\sum}_{i=1}^n\frac{(|r_i -m_i|)^pg(b_i)^{I(r_i\in\mathcal{E}_i)}}{f(w_i)}$$ where $r_i$ is the $i-$th ordered residual and $g(b_i)$ is a function of the distance of the residual point to the boundary of the envelope: $$b_i=
\begin{cases}
r_i-u_i, \mbox{ if } r_i>u_i\\
l_i-r_i, \mbox{ if } r_i<l_i
\end{cases}
$$ The indicator function $I(r_i\in\mathcal{E}_i)$ is equal to 1 if the residual point is contained in the envelope and equal to 0 otherwise, therefore the penalty function $g$ only influences the metric if the point is outside of the envelope. The variable $p$ can take two different values depending on the value of $p$. When the value of $p = 1$, the $L_1$ norm is considered and for $p = 2 $, the squared Euclidean distance is considered. The function $f(w)$ is the function for envelope width and acts as a scaling factor and has been tested for three different variations:

\begin{itemize}
    \item no scaling:  $f(w) = 1$
    \item inverse linear scaling: $f(w) = w$
    \item squared inverse scaling: $g(w) = w^2$
\end{itemize}

And we tested five different variations of $g(b)$:

\begin{itemize}
\item constant/no penalty: $g(b) = 1$,
\item unlimited linear increase: $g(b)=\displaystyle\alpha + \gamma b$,
\item saturated increase (ratio): $g(b)= \displaystyle\frac{\alpha + \gamma_1 b}{1+\gamma_2 b}$,
\item saturated increase (logistic): $g(b)=\displaystyle\frac{\alpha + \gamma}{1 + \exp{-\delta + (b - \eta)}}$ and 
\item saturated increase (hyperbolic tangent): $g(b)=\displaystyle\alpha + \gamma\tanh{\delta b}$.
\end{itemize}

The hyper-parameters $\alpha$, $\beta$, $\gamma$, $\gamma_1$, $\gamma_2$, $\eta$ and $\delta$ are assumed to be known and fixed. The penalties are introduced to differentiate, for instance, residual points that are close to either $u_i$ or $l_i$, but inside the envelope (and therefore expected under the fitted model), from points barely outside of the envelope, which should be more penalised, since that would not be expected under the fitted model most of the time.

We carried out a simulation study with 1,000 simulated samples from each of three sample sizes (20, 50, and 100) and three parent models (Poisson, negative binomial with a quadratic variance function with strong and mild overdispersion, negative binomial with a linear variance function with strong and mild overdispersion). We fitted three models to each simulated sample (Poisson and negative binomial with quadratic and linear variance functions), produced a half-normal plot with a simulated envelope for the Pearson residuals and computed $d_i$.

\subsection{Results: Appendix}
\begin{figure}[htb]
\centering
\includegraphics[width= 0.95\textwidth]{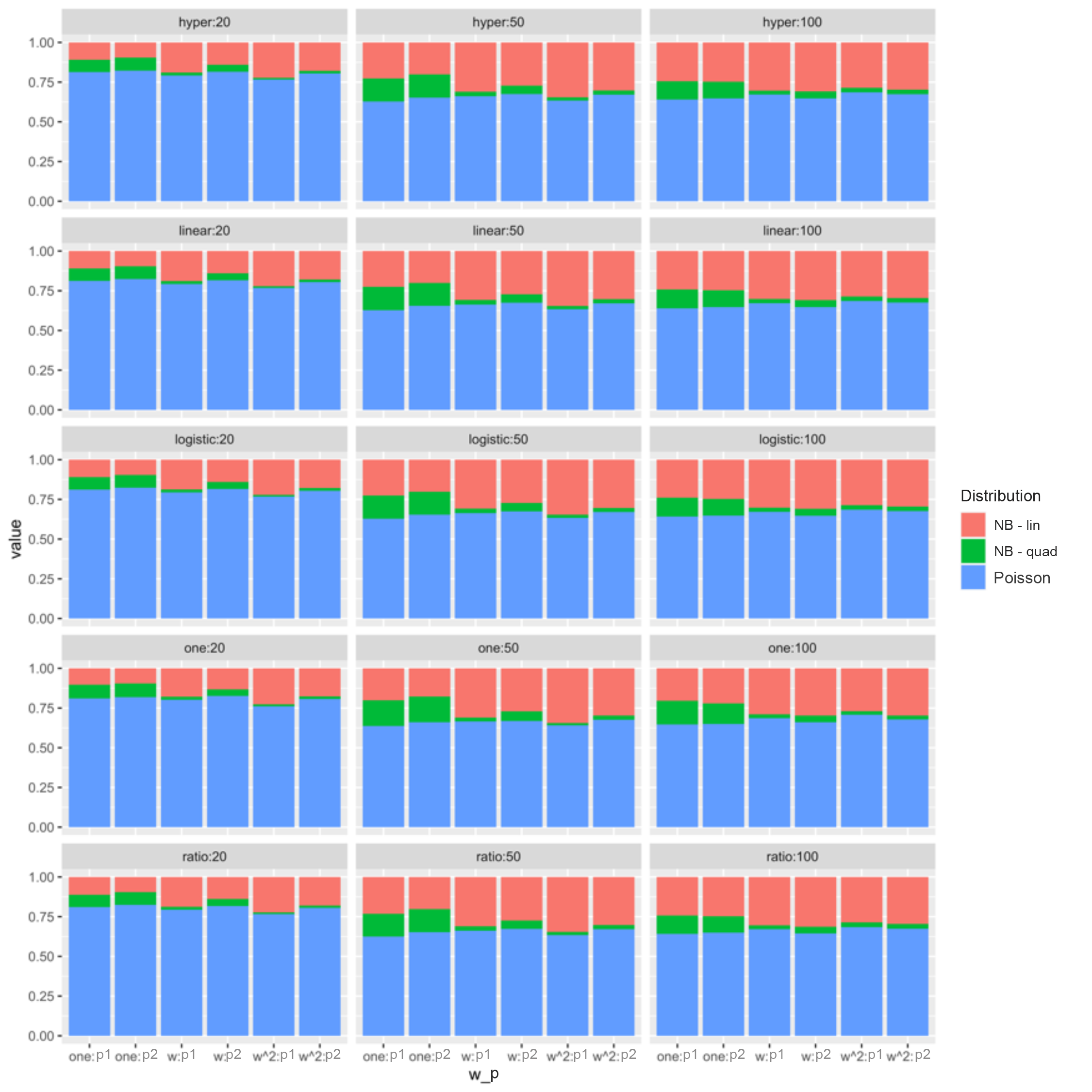}
\caption{Figure generated when the parent model is the Poisson}\label{app:poisson}
\end{figure}

\begin{figure}[htb]
\centering
\includegraphics[width= 0.95\textwidth]{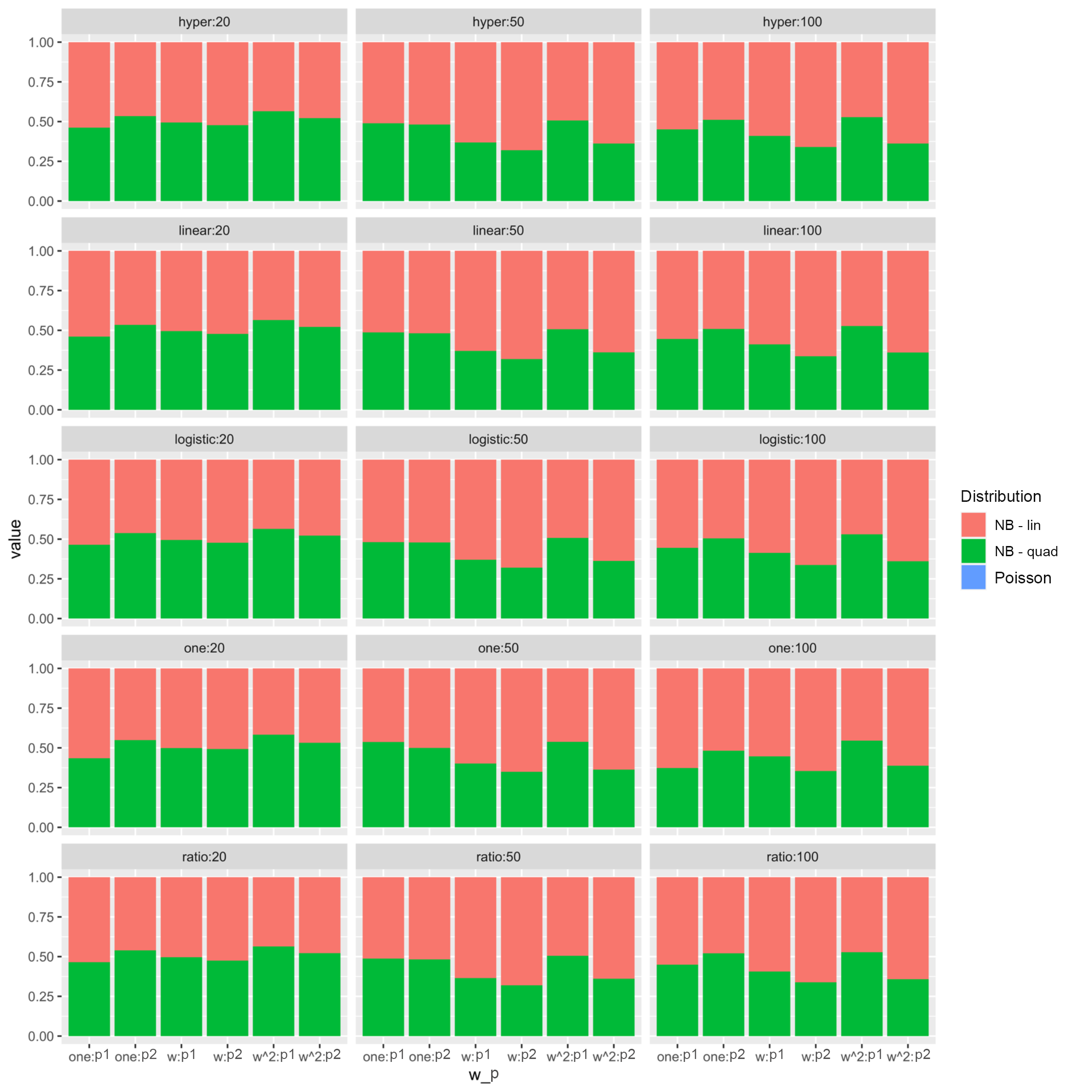}
\caption{Figure generated when the parent model is the NB-quad with a dispersion of $2$}\label{app:negbinm}
\end{figure}

\begin{figure}[htb]
\centering
\includegraphics[width= 0.95\textwidth]{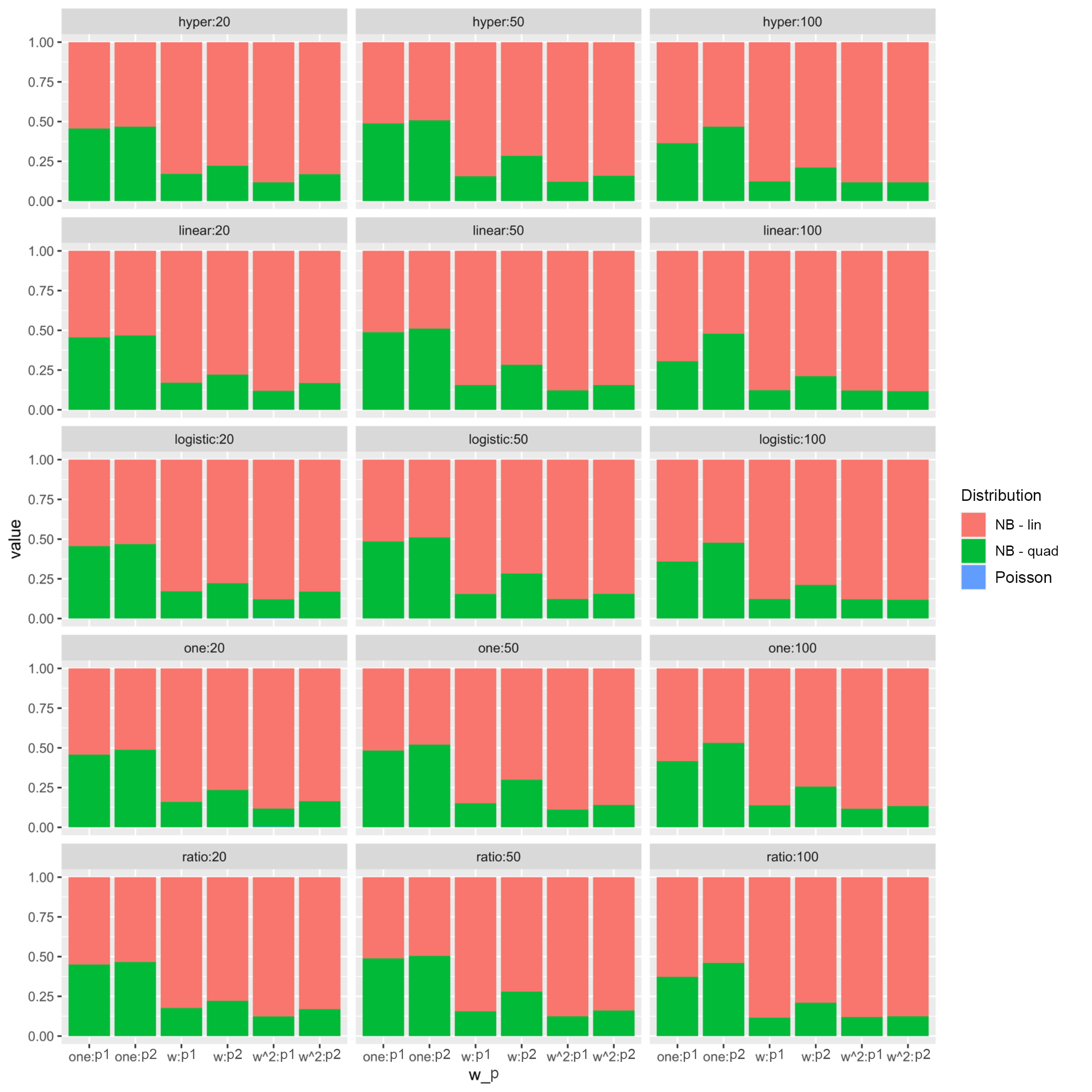}
\caption{Figure generated when the parent model is the NB-lin with a dispersion of $5$}\label{app:negbinlinh}
\end{figure}

The results are shown as three barplots with each barplot corresponding to each parent model. Each row in the bar plot shows the  each of the $g(b)'s$ considered and each column denotes the sample sizes considered $(20, 50, 100)$. Each bar in the individual block corresponds to the combination of type of distance ($p=1$: $L_1$ norm, $p=2$: squared Euclidean distance) and $f(w)'s$ considered and the $y$ axis shows the log transformed sum of the distance values for each fitted model. The fitted models are given in the legend; the distance metric values were calculated for NB-lin, NB-quad and Poisson as parent models, respectively. It is clearly evident from all barplots (Figure~\ref{app:poisson}, Figure~\ref{app:negbinm} and Figure~\ref{app:negbinlinh}) there is no difference between the penalty functions ($g(b)$) in terms of performance. There is negligible effect of scaling factor on the efficiency of the distance metric as it does not provide an  added ability in selecting the best performing model. The reasons for the null performance of added factors is presumed to be because we are recreating perfect scenarios where parent models are fitted to data generated by themselves or by closely related models. The poor performance of $g(b)$ might also be attributed to the small hyper-parameter values and since all the scenarios are perfect there are only few residuals falling outside the envelope.

\end{document}